# Bug shallowness in open-source, Macintosh software

G Gordon Worley III
ge011023@pegasus.cc.ucf.edu

**Abstract**
*Central to the power of open-source software is bug shallowness, the relative ease of finding and fixing bugs. The open-source movement began with Unix software, so many users were also programmers capable of finding and fixing bugs given the source code. But as the open-source movement reaches the Macintosh platform, bugs may not be shallow because few Macintosh users are programmers. Based on reports from open-source developers, I, however, conclude that that bugs are as shallow in open-source, Macintosh software as in any other open-source software.*

## Introduction

In <u>The Cathedral and The Bazaar</u>, Raymond proposes the bazaar model of open-source development, summarized by several heuristics given throughout the essay [Raymond00]. Most of them, in my opinion, work well for all open-source software. However, two of them may present a problem for open-source, Macintosh software:

6. Treating your users as co-developers is your least-hassle route to rapid code improvement and effective debugging.
7. Release early. Release often. And listen to your customers.

Which leads to:

8. Given a large enough beta-tester and co-developer base, almost every problem will be characterized quickly and the fix obvious to someone.

In other words, **(6)** and **(7)** make bugs shallow **(8)** [Raymond00].

In this paper I will explain why I believe these heuristics may fail in open-source, Macintosh software to cause a reduction in bug shallowness and will show how well my beliefs correspond with real-world development.

## Raymond's cathedrals and bazaars in relation to bug shallowness

Raymond uses the terms cathedral and bazaar as metaphors for models of open-source development. In the cathedral model, a few developers do all the work and, although the source code is published, there is little expectation (or even active discouragement) of community programming support. Emacs, BSD, and Darwin are examples of cathedral development. The bazaar model, in contrast, has a small team of administrators (usually just one person) who coordinates a large community of co-developers. The best known example is Linux, although because of Raymond, the bazaar model is, if not achieved, strived for by most open-source projects. [Raymond00]

Further, referring back to **(8)**, bug shallowness is a term Raymond and Torvalds use to describe how easily bugs are fixed. Although bug shallowness is not presented as a rigorous, technical term, its metaphor is simple to follow: bugs are `buried' in programs; shallow bugs can be `dug' out quickly and fixed; deep bugs take a long time to be `dug' out. The point

Raymond makes is that, in the bazaar model, most bugs become shallow by virtue of having many `diggers' who can find and fix bugs. Thus we can make bug shallowness technical by defining it as the average time between bug introduction and bug correction. The bazaar model's benefits hinge on that time being shorter than the cathedral model's.[1]

To get these improvements in bug shallowness, the bazaar model depends on community programming support. Without it, as may be the case on the Macintosh, the bazaar model will not operate.

**The trouble with co-developers and finicky users in the bazaar model on the Macintosh**

Macintosh users appear unique among open-source software users. Unix `users' are expected to have some technical knowledge of their operating system and, possibly, computer programming ability. Mac users, though, have an expectation that they *won't* need technical knowledge or programming ability to use software [Lewis00]. For example, on the popular software site VersionTracker, negative reviews of Macintosh software often cite complaints that they have to know too much to make the program work as expected [VersionTracker04]. Of course, all computer users, regardless of operating system, would, all else being equal, prefer software that required less knowledge to make it function, but Macintosh users seem actual opposed to software that requires technical knowledge, even when there are no alternatives. This suggests that the key prerequisite for **(6)**, the presence of technically inclined and, preferably, programmer users, does not exist in the Macintosh user base.

Windows has a large user base and a large developer base. As we might expect, then, there is a lot of open-source software for Windows. When it comes to **(7)**, Windows software is at an advantage because Windows users at more accustomed to the presence of buggy software since, even if the *relative* amount of poor software is equal on all platforms, the *total* amount of poor software for Windows is greater than on any other platform. Macintosh users are, traditionally, at the other end, having a small user base and an even smaller developer base, and thus comparatively little software, leading to less *total* buggy software and less tolerance for buggy software. This difference leads to higher expectations of quality and rejection of even gratis software if it is too difficult to use. A good example of this is the reaction to Apple's Mac OS X Public Beta. Many Macintosh users opined against Apple for releasing buggy software, reckoning it better that Apple keep software to themselves until done [Siracusa00]. This antibug attitude suggests that any open-source, Macintosh software project will only be successful (have users) if it is at least of the minimum reliability standards expected by Macintosh users.

Although Raymond and others mention the importance of initial development, in the case of Macintosh software this eliminates much of the chance for **(7)** to work because most of the bugs must have been squashed before users will use the software [Raymond00] [Kesteloot98] [Bezroukov99].

Thus, I hypothesize that, without many co-developers and without users willing to use buggy software, one of the primary purported benefits to the bazaar model is lost on Macintosh software: bugs will be no more shallow than they are in cathedral open-source

---

[1] This paragraph was added to the paper after its initial presentation in response to reader questions.

development models or, possibly, even closed-source development models[2].

**Survey of open-source developers**

To assess the validity of my hypothesis, I conducted a qualitative survey of open-source developers. In mid-July, I used the random project link on freshmeat.net to e-mail 100 listed project authors the survey in Appendix A. At the end of two weeks, I received 17 responses, 1 refusal, and 17 bounces (64 no responses). Recipients were asked the following questions:

1. What is your name and the name of your project?

2. What OSes or environments is your project designed to run on?

3. How do your users respond to buggy releases? Do they help you try to find the problems? Do they complain about the release of buggy software?

4. Do non-developer users report bugs to you? If so, how often? How do you respond?

5. How many active software developers do you have? How adequately would you say this meets with the project's needs?

6. How technically adept are your users? Are they novices? Do they have technical knowledge of their computers? Can they program?

The first question was used to assert the uniqueness of responses only and the second question was used to separate Macintosh projects from others. The remainder of the questions were phrased to try to discover the projects' bug shallowness, hopefully without biasing towards or away from my hypothesis.

## *Results*

Of the 17 responses, 13 projects claimed to run on any POSIX-like system (including Mac OS X) and 7 of these explicitly named Mac OS X as a supported platform. Of this, 4 projects had Mac OS X only or Mac OS X specific software, suggesting Mac OS X is an important operating system in the eyes of open-source developers (no one specifically mentioned `classic' versions of the Mac OS).

Support of Mac OS X did not correlate with any indicators of bug shallowness. Number of co-developers, frequency of bug reports, technical adeptness of users, and number of patches (when reported) all varied independently of reported operating system or environment in the survey sample.

## *Discussion*

From the evidence, I must reject my hypothesis and conclude that support of Mac OS X has no effect on bug shallowness. This is in itself, though, an important result, because it shows us that the bazaar model will work for projects that support Mac OS X, even for projects that run only on Mac OS X. This is an especially important piece of evidence for open-source proponents who face opposition from those who claim "that may work on *other* operating systems, but it won't work on *mine* with *my* users".

The most obvious reason my hypothesis may have failed was that the sample was inadequate because it was too small and vague. This survey employed qualitative methods, in part because I wasn't entirely sure what kind of responses developers would want to give and because I do not believe directly asking for a quantitative

---

[2] If there are no co-developers, it is, in fact, *exactly* like closed-source development from a bug correction and feature addition perspective

assessment of bug shallowness in a project would be meaningful because it is too complex a variable to guess without sufficient and careful prompting. I now, though, have a better idea of what developers know about their projects, so a good follow-up to this paper would be to perform a new survey with both more developers and quantitative questions.

My hypothesis may also have failed because my reasoning is no longer valid. Most of my assumptions come from the Macintosh tradition *prior* to Mac OS X. However, Mac OS X has, no doubt, altered Macintosh user culture, perhaps to render my assumptions false. For example, Mac OS X has seen an influx of Unix users who want a nicer graphical interface. And, as now seems likely to me, the antibug belief may never have been—an idea purported by a few outspoken individuals but rarely practiced by most users. But these are only my most likely candidates; there are doubtless many more reasons (and perhaps the correct ones) that my assumptions may have failed.

## Conclusion

Beyond the decision to open-source or not, Macintosh developers should feel confident that the bazaar model can work for open-source, Macintosh software. The key word, though, is `can'. Macintosh, open-source developers face the same problems other open-source developers face—coordinating developers and co-developers, attracting users, managing the project, etc.. Macintosh, open-source developers just don't have any extra difficulties from bug shallowness. Although this may seem a minor result, it's an important incremental step in understanding the dynamics of open-source, Macintosh software.

## Appendix A: Developer survey

Hi. My name is Gordon Worley and I'm contacting you because you were listed as the primary developer of a software project selected using the `random project' link on freshmeat.net. I'm writing a paper about models of open-source development and, if you have a few minutes, would greatly appreciate your answers the following questions:

1. What is your name and the name of your project?

2. What OSes or environments is your project designed to run on?

3. How do your users respond to buggy releases? Do they help you try to find the problems? Do they complain about the release of buggy software?

4. Do non-developer users report bugs to you?  If so, how often?  How do you respond?

5. How many active software developers do you have?  How adequately would you say this meets with the project's needs?

6. How technically adept are your users?  Are they novices?  Do they have technical knowledge of their computers?  Can they program?

Please note that all data collected will be used in aggregate and specific answers will not be tied to your name or the name of your project (I ask for your name and the name of your project for record keeping purposes only).

I'm sorry for not taking the time to customize this message, but I'm trying to survey many developers in a relatively short period of time.  Please respond within two weeks.  Thank you for your assistance.  If you so desire, I will contact you when the paper is published and send you a link to a free, electronic version.